%% file: main.tex
\documentclass[sigconf]{acmart}

\acmConference[MSR 2024]{21st International Conference on Mining Software Repositories}{April 2024}{Lisbon, Portugal}
\AtBeginDocument{%
  \providecommand\BibTeX{{%
    \normalfont B\kern-0.5em{\scshape i\kern-0.25em b}\kern-0.8em\TeX}}}

\setcopyright{acmcopyright}
\copyrightyear{2018}
\acmYear{2018}
\acmDOI{XXXXXXX.XXXXXXX}

\usepackage{colortbl}
\usepackage{enumitem}

\begin{document}

\title{Incivility in Open Source Projects: A Comprehensive Annotated Dataset of Locked GitHub Issue Threads}

\author{Ramtin Ehsani}
\affiliation{%
  \institution{Drexel University}
  \city{Philadelphia, PA}
  \country{USA}}
\email{ramtin.ehsani@drexel.edu}

\author{Mia Mohammad Imran}
\affiliation{%
\institution{Virginia Commonwealth University}
  \city{Richmond, VA}
  \country{USA}}
\email{imranm3@vcu.edu}

\author{Robert Zita}
\affiliation{%
\institution{Elmhurst University}
  \city{Elmhurst, IL}
  \country{USA}}
\email{rzita8729@365.elmhurst.edu}

\author{Kostadin Damevski}
\affiliation{%
\institution{Virginia Commonwealth University}
  \city{Richmond, VA}
  \country{USA}}
\email{kdamevski@vcu.edu}

\author{Preetha Chatterjee}
\affiliation{%
  \institution{Drexel University}
  \city{Philadelphia, PA}
  \country{USA}}
\email{preetha.chatterjee@drexel.edu}


\begin{abstract}
In the dynamic landscape of open source software (OSS) development, understanding and addressing incivility within issue discussions is crucial for fostering healthy and productive collaborations. This paper presents a curated dataset of 404 locked GitHub issue discussion threads and 5961 individual comments, collected from 213 OSS projects. We annotated the comments with various categories of incivility using Tone Bearing Discussion Features (TBDFs), and, for each issue thread, we annotated the triggers, targets, and consequences of incivility. We observed that \textit{Bitter frustration}, \textit{Impatience}, and \textit{Mocking} are the most prevalent TBDFs exhibited in our dataset. The most common triggers, targets, and consequences of incivility include \textit{Failed use of tool/code or error messages}, \textit{People}, and \textit{Discontinued further discussion}, respectively. This dataset can serve as a valuable resource for analyzing incivility in OSS and improving automated tools to detect and mitigate such behavior.
\end{abstract}


\maketitle

\input{sections/intro}
\input{sections/data_extraction}
\input{sections/data_description}
\input{sections/research_opportunities}
\input{sections/related_works}

\bibliographystyle{ACM-Reference-Format}
\bibliography{msr, career}

\end{document}

%% file: sections/intro.tex
\section{Introduction}
Issue trackers are pivotal in open source software (OSS) projects, facilitating effective monitoring, organization, and management of work~\citep{github_issue}. They serve as a central hub for diverse user and developer feedback, spanning ideas, tasks, features, and bug reports~\citep{issueTracking2019}. Within issue discussion threads, interactions range from constructive exchanges to unhealthy, uncivil, and toxic behaviors, which can hinder developer participation and productivity~\citep{Ferreira_2022_how_heated, millerDidYou, Gachechiladze_anger}.
Existing research has consistently highlighted the adverse impacts of toxicity and incivility within collaborative spaces, with consequences ranging from project abandonment to lower contribution rates ~\citep{SarkerToxiCR, millerDidYou, EgelmanPredict2020}, especially impacting people from underrepresented communities ~\citep{patchesGender2012, Diversity_Crisis2021, Trinkenreich2022}. Furthermore, these kinds of negative interactions can even affect developers' mental health, resulting in conditions such as stress and burnout ~\citep{RamanStress2020}.

Understanding negative interactions in software projects has gained significant attention in recent years~\cite{FerreiraSTFU2021, SarkerToxiCR, RamanStress2020, EhsaniMorality}. For instance, Raman et al. developed an automated tool to detect toxicity in GitHub issue threads~\cite{RamanStress2020}. Building on Raman et al.'s work, Sarker et al. created a tool for categorizing GitHub and Gitter code review messages as toxic or non-toxic~\cite{SarkerToxiCR}. Complementing these automated approaches, Miller et al. conducted a qualitative analysis of 100 GitHub locked issue threads~\cite{millerDidYou}. In contrast to established categories of toxicity in other domains~\cite{BlauWorkplaceIncivility, rasool_how_2021, anjum_empirical_2018, MolinaCivilityFacebook, CyberbullyingGames2015, cyberDetection2015}, they identified nuanced and distinct causes of toxicity in OSS, such as \textit{entitlement} and \textit{arrogance}. In exploring the broader category of incivility, Ferreira et al. analyzed Linux mailing lists and GitHub issue threads~\cite{FerreiraSTFU2021, Ferreira_2022_how_heated} to study Tone Bearing Discussion Features (TBDFs) -- conversational characteristics demonstrated in a written sentence that convey a mood or style of expression.

In spite of the prior work in this area, there is still a lack of a robust and comprehensive approach to address uncivil interactions in OSS. Three key factors strongly contribute to this deficiency: (a) the scarcity of large annotated software engineering-specific datasets~\cite{ferreiraIncivility2022, imran2022data}, (b) a lack of deep understanding of the nuances of negative behavior in OSS (e.g., triggers, targets, and consequences) ~\cite{millerDidYou}, and (c) a lack of a comprehensive taxonomy consolidating different types of uncivil behaviors in SE. 
%
%
In this paper, we annotate and publish an incivility dataset of 404 locked GitHub issue threads (5,961 issue comments) in open-source repositories. 
To curate the dataset, we gathered issue threads from 213 projects on GitHub, which had at least 50 contributors. We gathered issues that were either explicitly labeled and locked as "too heated" or demonstrated clear characteristics indicative of heated discussions. After collecting this initial dataset, we manually analyzed and annotated the following attributes of incivility in open source: types of incivility, its triggers, associated targets, and consequences or aftermath of incivility as witnessed within these conversations. To support the high-quality annotation of these various attributes related to incivility, we developed an annotation tool using Streamlit~\cite{streamlit}, a Python library, to streamline the process for the annotators involved in this study.

Our annotated dataset reveals prevalent forms of incivility within OSS projects, with \textit{Bitter frustration}, \textit{Impatience}, and \textit{Mocking} emerging as the most recurrent types. Additionally, the most frequent triggers, targets, and consequences are \textit{Failed use of code}, \textit{People}, and \textit{Discontinued further discussion}, respectively. This nuanced understanding, derived from our dataset, provides a valuable foundation for future research to delve deeper into the intricacies of incivility within OSS communications. Analyzing its unique nature can pave the way for the development of targeted mitigation and detection tools, fostering more inclusive and collaborative environments within these communities. Our dataset is available on GitHub: 
\color{blue}\url{https://github.com/vcu-swim-lab/incivility-dataset}\color{black}.

%% file: sections/data_extraction.tex
\section{Methodology}

\input{sections/bigtable}

In this section, we detail our approach for curating a OSS incivility dataset. Our decision to focus on incivility rather than toxicity is driven by a deliberate choice. While these two concepts overlap, toxicity primarily involves language that harms others. Incivility, on the other hand, has a broader scope, encompassing issues that can disrupt constructive and technical discussions~\cite{FerreiraSTFU2021}. As highlighted by Sadeque et al.~\cite{sadeque-etal-2019-incivility}, the development of a fine-grained incivility detection tool presents a more intricate challenge compared to toxicity detection. We aim to provide a more nuanced understanding of negative discourse dynamics within the context of OSS projects.

\subsection{Data Collection}
In an effort to mitigate unproductive discussions, GitHub offers functionality that enables project maintainers to lock issue threads, thus preventing further discussion. During the locking process, maintainers can, but are not required to, label the reason the discussion was locked, e.g., as ``too-heated." These moderation tools serve the dual purpose of aiding project contributors in effectively managing and moderating discussions, while also intervening to stop overly contentious conversations when necessary. Locked and labeled discussions within OSS projects offer valuable research data. Since these labels are typically assigned by project maintainers, they serve as a reliable means to pinpoint specific conversation instances of interest to researchers.

Given our goal of curating GitHub issue threads likely to exhibit incivility, particularly within the context of active OSS projects, we established the following data collection procedure:
\begin{itemize}[leftmargin=*]
    \item The GitHub project must have a minimum of 50 contributors.
    \item The GitHub issues must have been created in the last 10 years, i.e., between ``2013-04-07'' and ``2023-10-24''.
    \item The issues must be locked and labeled as ``too-heated", ``off-topic'' or ``spam''. In cases where the issue was locked but the reason (label) was not explicitly stated, we selected the issues that included the text  "code of conduct" or "marked as abusive"  within the discussion. Such terms have previously used as clear indicators of potentially toxic or incivil discussions~\cite{millerDidYou}.
\end{itemize}

Since it was not possible to directly retrieve all the locked issues with the chosen labels from GitHub, we adopted two approaches to ensure the collection of as many of these issues as possible:

\textbf{GitHub API.} Leveraging GitHub's official API~\cite{github_repos}, we systematically accessed publicly available repositories one-by-one to examine their issue threads. This process was constrained by the rate limiting mechanism of the GitHub API, and thus, given a limited time budget, we were only able to collect data from approximately 600k OSS projects.

\textbf{GitHub Archive.} We used the GH Archive~\cite{gharchive}, a project dedicated to the recording of the public GitHub timeline and making it readily available for rapid querying and analysis. We used the BigQuery interface and the latest archive (up to the end of 2022) to retrieve GitHub issues where the locked issue type is labeled as ``too-heated", ``spam'', or ``off-topic''. 

\subsection{Data Selection}

Locked issue threads labeled as ``too-heated'' are the most evident candidates for inclusion in our dataset, due to the prevalence of uncivil conversations within them. Since it is possible for project maintainers to mislabel issues~\cite{ferreiraHeat2022}, the first two authors examined the collected issues labeled as ``too-heated'' and removed the ones that were obviously mislabeled.

Our motivation to extend our selection to ``spam'' and ``off-topic'' locked issue threads arises from the limited number of ``too-heated'' threads on GitHub.  To select issue threads labeled as ``off-topic'' or ``spam'' that have potentially uncivil content, the first two authors examined a total of 422 issue threads from these two categories, manually assessing their content and resolution. This effort allowed us to filter out issue threads that were solely ``spam'' or ``off-topic'' but not uncivil.

In total, following the application of our selection criteria to the collected repositories and issue threads, merging duplicate instances between the results of the two approaches, and manually filtering the noisy conversations in our dataset, we successfully gathered 404 instances -- 338 labeled as ``too-heated'', 21 occurrences of ``spam'', and 33 labeled as ``off-topic'', and 12 instances of issue threads locked without specified reasons but containing the keywords "code of conduct" or "marked as abusive".

\subsection{Annotation Tool}
To simplify and streamline the annotation process, we designed a bespoke web annotation tool. We used Streamlit, an open-source app framework, that enables the creation of web apps using data scripts. Our annotation app includes a secure login page, allowing annotators to access the tool with their unique login ID. To aid annotators, we incorporated annotation instructions into the tool's interface, ensuring ready access to category definitions. The GitHub issues are securely stored in an SQLite database, with a subset distributed to each annotator.
The web app screenshots and source code are available in our dataset's repository.

\subsection{Incivility Categories}

For each issue thread in our dataset, we manually annotated four categories: \textit{type of incivility}, \textit{trigger}, \textit{target}, and \textit{consequence}. Each category was selected based on existing literature on harmful interactions across various domains such as social media~\citep{Chohan_twitter, Alsoubai_fwb, Pfeffer_twitter, Hee_cyber}, online gaming communities~\citep{Kwak_game, kouToxic2020}, and software engineering~\citep{EgelmanPredict2020, SarkerToxiCR, Ferreira_2022_how_heated, FerreiraSTFU2021, Gachechiladze_anger, millerDidYou, cohenContextualizing2021, CheriyanNorm, CheriyanTowards, SocialInclusion}. 
Using a combination of deductive and inductive coding methods, we refined the feature set for each category to enhance the quality of our annotations~\citep{deductive_Theophilus, bradley_qualitative_2007}. This involved an iterative process of qualitative analysis, merging similar features and eliminating those that were too general or irrelevant, until we achieved a comprehensive set of features for each annotation category.

\textbf{Types of Incivility.}
To annotate types of incivility in our dataset, we utilized Ferreira et al.'s framework~\citep{FerreiraSTFU2021} for identifying uncivil textual elements, known as Tone Bearing Discussion Features (TBDFs), initially introduced by Coe et al~\citep{Coe_uncivil}.
TBDFs include conversational attributes in written sentences that convey specific moods or expressive styles. According to Ferreira et al.~\citep{FerreiraSTFU2021}, these characteristics are categorized into four sets: positive, negative, neutral, and uncivil. We focused solely on the uncivil characteristics for annotation. We also incorporated elements from the works of Miller et al.~\citep{millerDidYou} and Sarker et al.~\citep{SarkerToxiCR} to further enhance our feature set's comprehensiveness. The complete set of annotation categories in our dataset, along with their definitions and examples, is listed in Table \ref{tab:uncivil_features}.

\textbf{Trigger.}
Our aim was to identify the underlying triggers of incivility within GitHub conversations. The categorizations for this annotation were drawn from the extensive research of Ferreira et al.~\citep{FerreiraSTFU2021}, which includes triggers like \textit{Communication breakdown}, \textit{Rejection}, and \textit{Violation of community conventions}, and Miller et al.~\citep{millerDidYou}, which includes triggers such as \textit{Failed use of tool/code or error messages}, \textit{Past interactions}, \textit{Politics/ideology}, and \textit{Technical disagreement}. We expanded these categories by introducing an additional category, \textit{Unprovoked}, to capture instances of incivility without a discernible trigger.

\textbf{Target.}
This annotation aimed to pinpoint the specific target of incivility in conversations. The categorizations were derived from Miller et al.~\citep{millerDidYou}, including \textit{People}, \textit{Code/tool}, \textit{Company/organization}, \textit{Self-directed}, and \textit{Undirected}.

\textbf{Consequence.}
The primary goal of this annotation was to uncover the repercussions of incivility as observed within conversations, thereby shedding light on the aftermath of such interactions in developer communications. These categorizations were inspired by existing research by Ferreira et al.~\citep{FerreiraSTFU2021}, which includes consequences such as \textit{ Discontinued further discussion}, \textit{Provided technical explanation}, \textit{Accepting criticism}, and \textit{Trying to stop the incivility}, and Miller et al.~\citep{millerDidYou}, which includes \textit{Invoke Code of Conduct}, \textit{Turning constructive}, and \textit{Escalating further}.

\subsection{Data Annotation Procedure}
A total of 19 university students (junior undergraduate=2, senior undergraduate=16, master's=1) studying computer science were recruited as annotators. These students were recruited through email outreach. After completing a consent form, clear annotation instructions, including examples, were provided in the form of an external document (with key parts of it repeated in the annotation tool as reminders).  All annotators had prior experience with GitHub (0-2 years=13, 2-4 years=5, 4+ years=1), and one annotator had prior experience contributing to open source projects.
Each student annotated approximately 20 issue threads.

To further improve the annotation's quality, we use GPT-4~\cite{openai2023gpt4} (via the `gpt-4' API), which has shown promising results in text annotation, in some cases outperforming humans~\cite{gilardi2023chatgpt, dai2023chataug, huang2023chatgpt}.
We formulate a prompt in which GPT-4 systematically evaluates student annotations utilizing a 5-point scale. Instances with an agreement score below 3 are considered disagreements in our analysis.
We found 549 out of 5,961 utterances where GPT-4 disagreed with the human annotation. Two of the authors of this paper, manually checked those 549 utterances to resolve the disagreements, revising the annotation of 311 instances. For example, the comment {\em``Calm down, Please.''} was initially labeled as {\em Impatience}. However, upon a manual review and considering the context of the ongoing conversation within the issue thread, the annotation was revised to {\em None}. This process ensured a high accuracy and reliability of the final annotated dataset.

%% file: sections/bigtable.tex
\begin{table*}
\caption{Uncivil Features; Sources are Color Coded for \colorbox{yellow}{Ferreira et al.}, \colorbox{cyan}{Sarker et al.}, and \colorbox{lime}{Miller et al.} (N=No. of Issues)}
\vspace{-0.3cm}
\centering
\scriptsize
\resizebox{\textwidth}{!}{%
\begin{tabular}{|l|l|l|l|l|l|}
\hline
\textbf{Feature} & \textbf{Definition and Example} & \textbf{\begin{tabular}[c]{@{}l@{}}Most Common Triggers\end{tabular}} & \textbf{\begin{tabular}[c]{@{}l@{}}Most Common Targets\end{tabular}} & \textbf{\begin{tabular}[c]{@{}l@{}}Most Common Consequences\end{tabular}} & \textbf{{\begin{tabular}[c]{@{}l@{}}\# of Issue\\Comments\end{tabular}}} \\ \hline

\cellcolor[HTML]{FFFE65}Bitter frustration & \textit{\begin{tabular}[c]{@{}l@{}}\textbf{Def.}  expressing strong frustration\\ \textbf{e.g.} Fixing clippy warnings isn't adding anything \\ for users\end{tabular}} & \begin{tabular}[c]{@{}l@{}}Failed use of tool/code or error messages (N=49),\\ Technical disagreement (N=49)\end{tabular} & \begin{tabular}[c]{@{}l@{}}People (N=107),\\ Code/tool (N=54)\end{tabular} & \begin{tabular}[c]{@{}l@{}}Discontinued further discussion (N=71),\\ Escalating further (N=63),\\ Provided technical explanation (N=37)\end{tabular} & \textbf{492} \\ \hline
\cellcolor[HTML]{FFFE65}Impatience & \textit{\begin{tabular}[c]{@{}l@{}}\textbf{Def.} express a feeling that it is taking too long\\ \textbf{e.g.} I am locking this thread. It is becoming useless\end{tabular}} & \begin{tabular}[c]{@{}l@{}}Technical disagreement (N=34),\\ Failed use of tool/code or error messages (N=29),\end{tabular} & \begin{tabular}[c]{@{}l@{}}People (N=70),\\ Code/tool (N=37)\end{tabular} & \begin{tabular}[c]{@{}l@{}}Discontinued further discussion (N=49),\\ Escalating further (N=36),\\ Provided technical explanation (N=31)\end{tabular} & 264 \\ \hline
\cellcolor[HTML]{FFFE65}Mocking & \textit{\begin{tabular}[c]{@{}l@{}}\textbf{Def.} making fun of someone else\\ \textbf{e.g.} congrats, you won an award for the best \\ support of the month\end{tabular}} & \begin{tabular}[c]{@{}l@{}}Failed use of tool/code or error messages (N=19),\\ Technical disagreement (N=15),\\ Communication breakdown (N=14)\end{tabular} & \begin{tabular}[c]{@{}l@{}}People (N=59),\\ Code/tool (N=12)\end{tabular} & \begin{tabular}[c]{@{}l@{}}Escalating further (N=37),\\ Discontinued further discussion (N=32)\\ Provided technical explanation (N=20)\end{tabular} & 180 \\ \hline
\cellcolor[HTML]{FFFE65}Irony & \textit{\begin{tabular}[c]{@{}l@{}}\textbf{Def.} signify the opposite in a mocking way\\ \textbf{e.g. }Ok, you win, have fun arguing forever instead \\ of proposing a solution\end{tabular}} & \begin{tabular}[c]{@{}l@{}}Technical disagreement (N=11),\\ Failed use of tool/code or error messages (N=10)\end{tabular} & \begin{tabular}[c]{@{}l@{}}People (N=20),\\ Code/tool (N=16)\end{tabular} & \begin{tabular}[c]{@{}l@{}}Escalating further (N=21),\\ Discontinued further discussion (N=19),\\ Provided technical explanation (N=12)\end{tabular} & 64 \\ \hline
\cellcolor[HTML]{FFFE65}Vulgarity & \textit{\begin{tabular}[c]{@{}l@{}}\textbf{Def.} using profanity or improper language\\ \textbf{e.g.} It honestly looks like they don't give a sh*t, \\ rules this out as an option for me!\end{tabular}}  & \begin{tabular}[c]{@{}l@{}}Failed use of tool/code or error messages (N=13),\\ Technical disagreement (N=9)\end{tabular} & \begin{tabular}[c]{@{}l@{}}People (N=31),\\ Code/tool (N=12)\end{tabular} & \begin{tabular}[c]{@{}l@{}}Escalating further (N=26),\\ Discontinued further discussion (N=16),\\ Trying to stop the incivility (N=12)\end{tabular} & 71 \\ \hline
\cellcolor[HTML]{FFFE65}Threat & \textit{\begin{tabular}[c]{@{}l@{}}\textbf{Def.} put a condition impacting the result of discussion\\ \textbf{e.g.} This is the final notice. Be honest, respectable, \\ and collaborative\end{tabular}} & \begin{tabular}[c]{@{}l@{}}Communication breakdown (N=4),\\ Violation of community conventions (N=3)\end{tabular} & \begin{tabular}[c]{@{}l@{}}People (N=10),\\ Code/tool (N=5)\end{tabular} & \begin{tabular}[c]{@{}l@{}}Discontinued further discussion (N=10),\\ Escalating further (N=8),\\ Trying to stop the incivility (N=7)\end{tabular} & 23 \\ \hline

\cellcolor[HTML]{bfff00}Entitlement & \textit{\begin{tabular}[c]{@{}l@{}}\textbf{Def.} expecting special privileges\\ \textbf{e.g.} Or you could start contributing instead of \\ bashing people who actually do the work\end{tabular}} & \begin{tabular}[c]{@{}l@{}}Technical disagreement (N=12),\\ Failed use of tool/code or error messages (N=9)\end{tabular} & \begin{tabular}[c]{@{}l@{}}People (N=30),\\ Code/tool (N=10)\end{tabular} & \begin{tabular}[c]{@{}l@{}}Escalating further (N=20),\\ Discontinued further discussion (N=20), \\ Provided technical explanation (N=12)\end{tabular} & 69 \\ \hline
\cellcolor[HTML]{34CDF9}Insulting & \textit{\begin{tabular}[c]{@{}l@{}}\textbf{Def.} remarks directed at another person\\ \textbf{e.g.} Seems like only thing you can do so far is talk,\\ come back when you will have any skill to show.\end{tabular}} & \begin{tabular}[c]{@{}l@{}}Failed use of tool/code or error messages (N=23),\\ Technical disagreement (N=19)\end{tabular} & \begin{tabular}[c]{@{}l@{}}People (N=57),\\ Code/tool (N=19)\end{tabular} & \begin{tabular}[c]{@{}l@{}}Escalating further (N=41),\\ Discontinued further discussion (N=33),\\ Provided technical explanation (N=17)\end{tabular} & 174 \\ \hline
\cellcolor[HTML]{34CDF9}\begin{tabular}[c]{@{}l@{}}Identity attacks/\\ Name-calling\end{tabular} & \textit{\begin{tabular}[c]{@{}l@{}}\textbf{Def.} Race, Religion, Nationality, Gender, Sexual- \\oriented attacks\\ \textbf{e.g.} I would not be surprised if this database is \\ maintained by the Russians\end{tabular}} & \begin{tabular}[c]{@{}l@{}}Politics/ideology (N=6)\\ Failed use of tool/code or error messages (N=5)\end{tabular} & \begin{tabular}[c]{@{}l@{}}People (N=15),\\ Company/organization (N=7)\end{tabular} & \begin{tabular}[c]{@{}l@{}}Escalating further (N=13),\\ Discontinued further discussion (N=8),\\ Invoke Code of Conduct (N=5)\end{tabular} & 28 \\ \hline
\end{tabular}%
}
\label{tab:uncivil_features}
\end{table*}

%% file: sections/data_description.tex
\section{Dataset Description}

\textbf{Incivility Annotations.} Of the 5,961 issue comments analyzed, 1,365 were annotated with an incivility feature. The distribution of these annotations is detailed in Table \ref{tab:uncivil_features}. \textit{Bitter frustration}, \textit{Impatience}, and \textit{Mocking} are the most recurrent uncivil features in this dataset. Among the 404 issue threads, 319 have at least one uncivil feature annotation. Of the 85 threads without any identified uncivil feature, 78 were locked as ``too heated," 2 as ``spam," 2 without specified reasoning, and 1 as ``off-topic."

\begin{figure}[t]
    \centerline{\includegraphics[scale=0.27]{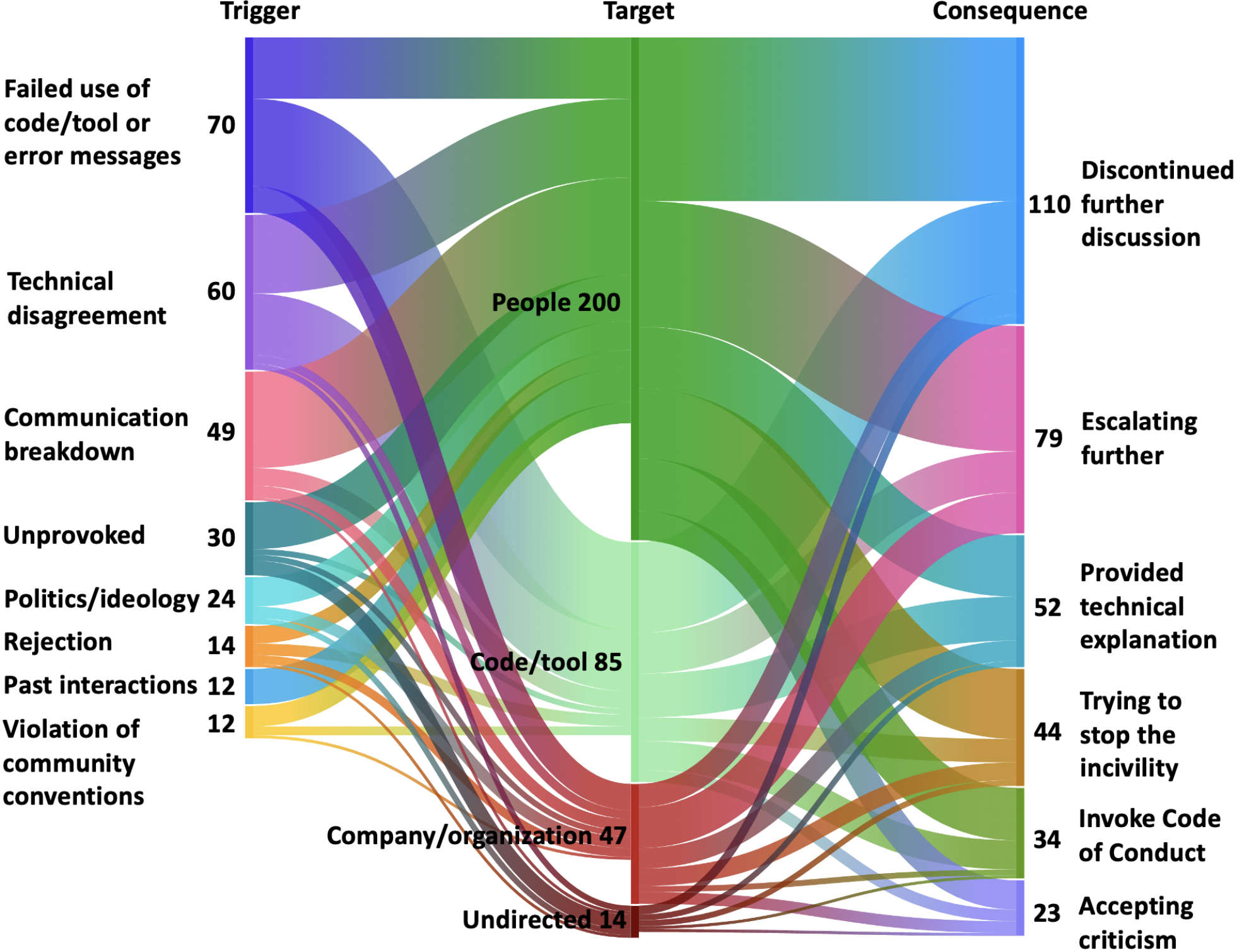}}
    \caption{Triggers, Targets, and Consequences of Incivility}
    \label{fig:sankey}
\end{figure}

\textbf{Issue Thread Annotations.} The distribution of annotated triggers, targets, and consequences within this dataset is presented in Figure \ref{fig:sankey}. \textit{Failed use of tool/code or error messages}, \textit{Technical disagreement}, and \textit{Communication breakdown} are the most prevalent triggers. The most frequent targets are \textit{People} and \textit{Code/tool}, while the most common consequences include \textit{Discontinued further discussion}, \textit{Escalating further}, and \textit{Provided technical explanation}. Notably, none of the uncivil issue threads in this dataset transitioned into constructive discussions (\textit{Turning Constructive N=0}). Annotators could label each issue with multiple consequences, with combinations like \textit{[Escalating further, Discontinued further discussion]}, \textit{[Invoke Code of Conduct, Discontinued further discussion]}, and \textit{[Escalating further, Trying to stop the incivility]} being prominent.

\textbf{Observations.}
When uncivil discussions target people, the main triggers are \textit{Communication breakdown (N=33)}, \textit{Technical disagreement (N=27)}, and \textit{Failed use of tool/code or error messages (N=21)}, with the most common consequences being \textit{Discontinued further discussion (N=62)}, \textit{Escalating further (N=51)}, and \textit{Trying to stop the incivility (N=27)}. In contrast, when incivility targets Code/tool, the primary triggers are \textit{Failed use of tool/code or error messages (N=32)}, \textit{Technical disagreement (N=22)}, and \textit{Communication breakdown (N=6)}, with the most frequent consequences being \textit{Discontinued further discussion (N=34)}, \textit{Provided technical explanation (N=15)}, and \textit{Escalating further (N=14)}. An interesting finding is that \textit{Failed use of tool/code or error messages} as a trigger often leads to incivility directed at \textit{Code/tool}, whereas \textit{Technical disagreement} usually results in incivility aimed at \textit{People}. Figure \ref{fig:sankey} illustrates detailed relationships between targets, triggers, and consequences in this dataset, such as \textit{Communication breakdown} typically targeting \textit{People} and leading to the discontinuation of further discussion.

%% file: sections/research_opportunities.tex
\section{Research Opportunities}
Our dataset presents numerous opportunities for addressing and exploring challenges in sustainable software projects and developer productivity. The prevalence of toxic interactions and uncivil language within OSS communities has become a pressing issue, leading to negative emotional experiences and developer isolation. This dataset is a valuable resource for conducting comprehensive analyses of incivility within developer communications.

It offers the potential to train and refine automatic incivility detection tools. These tools can identify uncivil conversations and help mitigate disruptive interactions within discussions. Our annotations provide more than just flags for uncivil comments; they offer insights into the specific types of incivility present. This can be used to analyze developer interactions, highlighting the prevalence and nuances of different incivility types within developer communications. Previous research indicates that tools trained in other domains, like the Google Perspective API, are ineffective for software engineering (SE) corpora due to the unique nature of SE text~\cite{RamanStress2020, Sarker2020ABS}. Thus, developing an SE-specific incivility detection tool that understands the nuances of developer conversations, including SE jargon, is a valuable contribution to both the literature and the OSS community.

We focused our analysis on popular OSS projects on GitHub with significant contributor numbers. This approach allows us to examine the dynamics of incivility within these projects, identifying primary factors contributing to such occurrences, especially analyzing triggers, targets, and consequences of uncivil conversations in OSS.

Furthermore, our dataset enables exploration of how incivility might impact key project attributes and overall project health, including code quality and commit frequency. By employing triangulation studies or integrating data from GitHub's version control, we can assess the effects of incivility on developers' code quality and commits, providing a deeper understanding of team dynamics.

Code of Conduct is often used in OSS moderation. The dataset could also further enable analysis of moderation strategies and policies adopted by different open source projects to handle incivility. 

This dataset may help to forecast when a conversation is going to derail. The triggers and targets annotated in the dataset provide a foundation to explore personalized intervention approaches when automated tools detect potential early signs of uncivil conversations arising.

Additionally, considering the ongoing challenge of underrepresentation in OSS development, our dataset offers a unique opportunity to investigate how incivility affects individuals from underrepresented communities. By incorporating considerations of gender, race, and cultural aspects, and given the substantial populations of these projects, we can explore the implications of incivility on these communities.